\documentclass[%
 aip,
 sd,%
 amsmath,amssymb, 
 reprint,%
]{revtex4-1}

\usepackage{graphicx}
\usepackage{dcolumn}
\usepackage{bm}
\usepackage{color}

\def\eps{\varepsilon}

\def\be{\begin{equation}}
\def\ee{\end{equation}}
\def\ba{\begin{align}}
\def\bm{\begin{multline}}
\def\bfig{\begin{figure}[htb]}
\def\efig{\end{figure}}

\DeclareMathSymbol{\leqslant}{\mathalpha}{AMSa}{"36}
\DeclareMathSymbol{\geqslant}{\mathalpha}{AMSa}{"3E}
\DeclareMathSymbol{\doteqdot}{\mathalpha}{AMSa}{"2B}
\DeclareMathSymbol{\circlearrowright}{\mathalpha}{AMSa}{"08}
\DeclareMathSymbol{\subsetneq}{\mathalpha}{AMSb}{"28}
\DeclareMathSymbol{\supsetneq}{\mathalpha}{AMSb}{"29}
\renewcommand{\leq}{\;\leqslant\;}
\renewcommand{\geq}{\;\geqslant\;}

\newcommand{\dd}{{\rm d}}
\newcommand{\e}[1]{\,{\rm e}^{#1}\,}
\newcommand{\ii}{{\rm i}}

\newcommand{\upchi}{\raise 2pt \hbox{$\chi$}}

\newcommand{\caO}{{\mathcal O}}

\newcommand{\bbC}{{\mathbb C}}

\newcommand{\bsE}{{\boldsymbol E}}

\newcommand{\bsI}{{\boldsymbol I}}

\graphicspath{{./Graphics/}}

\begin{document}

\preprint{AIP/123-QED}

\title[Superadiabatic wave packet dynamics]
{Wave packet dynamics in the optimal superadiabatic approximation}%

\author{V. Betz}
\affiliation{Fachbereich Mathematik, TU Darmstadt}
\author{B. D. Goddard}%
\affiliation{The School of Mathematics and Maxwell Institute for Mathematical Sciences, University of Edinburgh
}%
\author{U. Manthe}%
\affiliation{Fakult\"at f\"ur Chemie, Universit\"at Bielefeld
}%

\date{\today}

\begin{abstract}
We explain the concept of superadiabatic approximations and 
show how in the context of the Born-Oppenheimer approximation 
they lead to an explicit formula that can be used to predict 
transitions at avoided crossings. 
Based on this formula, we present a simple method for computing 
wave packet dynamics across avoided crossings. 
Only knowledge of the adiabatic electronic energy levels 
near the avoided crossing is required for the computation.
In particular, this means that no diabatization procedure is necessary, 
the adiabatic energy levels
can be computed on the fly, and they only need to be computed to higher 
accuracy when an avoided crossing is detected. 
We test the quality of our method on the paradigmatic example 
of photo-dissociation of NaI, finding very good agreement with
results of exact wave packet calculations.
%
\end{abstract}

\keywords{quantum dynamics, non-adiabatic transitions, superadiabatic theory, avoided crossing}
\maketitle

\section{Introduction}

Superadiabatic approximations were first introduced by 
Michael Berry \cite{Be90} in the context of 
a generalized Landau-Zener Hamiltonian. 
They can be viewed as iterative improvements to the adiabatic approximation, 
in the same spirit that higher order perturbation
expansion improves first order perturbation theory. In the work of Berry 
a semiclassical approximation was made, and the nuclei were assumed 
to move classically. 
An extension of the theory to the full Born-Oppenheimer approximation
has been done in recent years (see Refs \onlinecite{BGT09,BG1,BG2}). 
In this introduction we discuss the theory of 
superadiabatic approximations in the Born-Oppenheimer context 
on an intuitive level; 
mathematical details will be given later. 

To understand superadiabatic approximations, consider first the adiabatic one. 
In the adiabatic representation, the frame of reference at each point in space is 
adjusted so that the electronic Hamiltonian is diagonal. In a wave packet picture, 
the frame of reference thus `moves with the nuclei', and it depends 
only on the position of the nuclei. By this procedure, the adiabatic representation
achieves that, in most situations, a wave packet started on an adiabatic energy level
remains there to a very good approximation. The errors to this invariance property are 
described by the kinetic coupling element. 

The superadiabatic representations improve
on the adiabatic one by taking finer aspects (like, for example, the momentum) 
of the wave packet into account. The result is that the kinetic coupling 
from the adiabatic representation is transformed into a 
coupling depending on the second or higher derivatives of the wave packet, but is now of 
much smaller magnitude. In many situations, already the quality of the 
adiabatic approximation is sufficient for describing the wave packet dynamics, and then
there is no need for further improvements. In other cases, however, 
it is advantageous to go beyond the adiabatic approximation. 

One of those situations are avoided crossings of electronic energy levels. 
At an avoided crossing, the 
adiabatic derivative couplings become large (but do not diverge). 
Typically the relevant nuclear configurations are 
indicated by a very small, but 
finite, energetic distance of the corresponding adiabatic energy levels, 
whence the name `avoided crossing'. 
As a result, a small but not negligible part of the 
nuclear wave packet traveling through the avoided 
crossing will make a transition to the previously unoccupied 
adiabatic energy level. 

\begin{figure}
\caption{Snapshots of a wave packet $\psi_-$ 
appearing on the previously unoccupied adiabatic 
energy level during a transition at an avoided crossing. 
Squared absolute values of position and momentum representation are displayed.
In the first snapshots, the original wave packet approaches the transition
point. A (spurious) wave packet starts to build up on the other adiabatic level.
At t=178~fs, the incoming wave packet is on top of the avoided 
crossing, and 
the spurious transmitted wave packet has grown to its maximal size. In the remaining 
snapshots, the incoming wave packet travels away from the avoided crossing, and 
the spurious wave packet starts to die down, revealing the much smaller true
transmitted wave packet. In the final snapshot, the transition is over, 
and only the true transmitted wave packet remains.}
\label{fig:stueckelberg}
\includegraphics[width = \columnwidth]{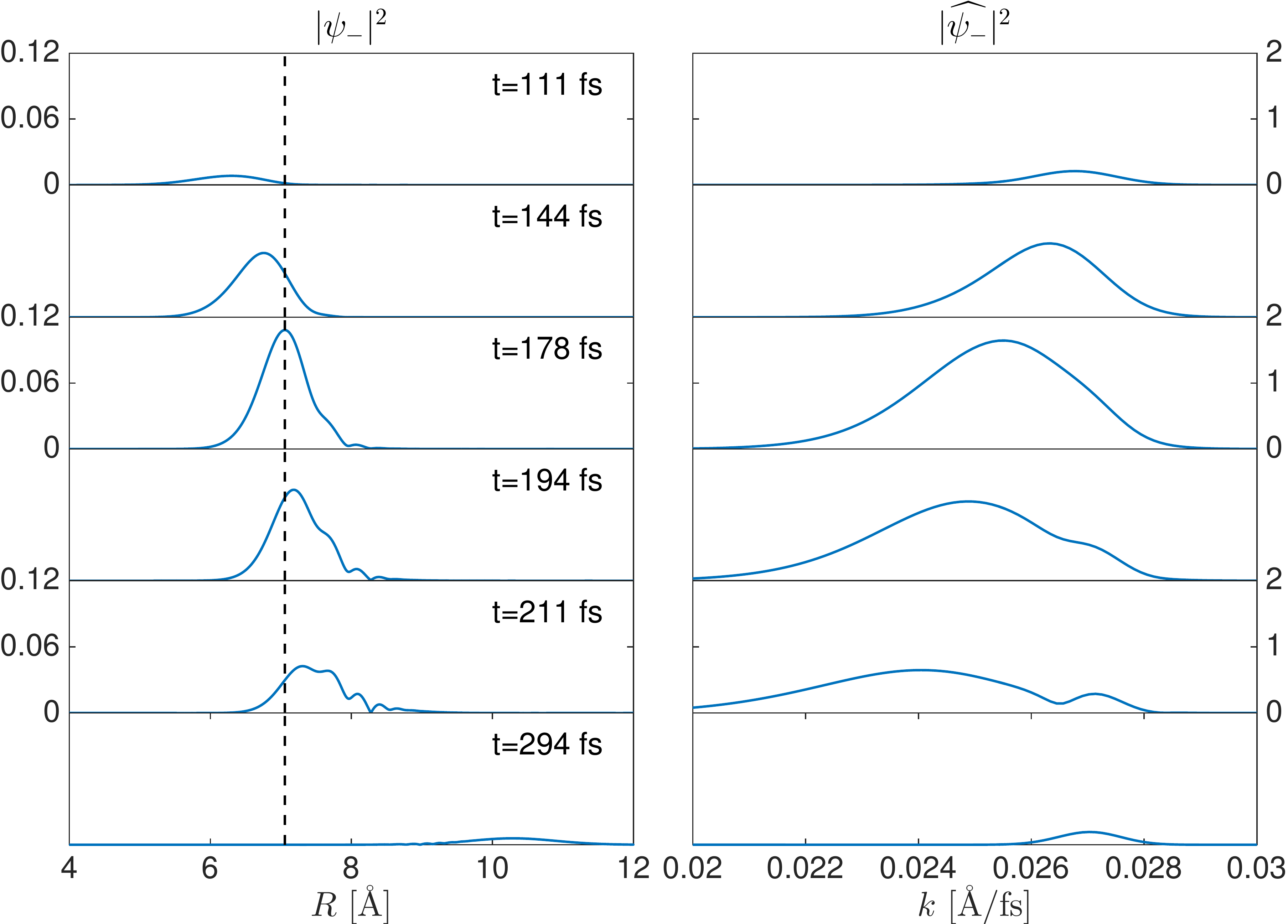}
\end{figure}

While such a transition can still be described in the adiabatic 
representation, this leads to apparently very complicated dynamics:
when the wave packet approaches the point of minimal 
separation of electronic energy levels, relatively large 
portions of it show up in the previously unoccupied adiabatic energy level, 
leading to St\"uckelberg oscillations. This is illustrated in 
Figure \ref{fig:stueckelberg}, which shows the exact 
adiabatic time evolution of the wave packet created at an avoided crossing
(at $R=7$\AA), for a slightly modified version of the NaI potential presented
in Section \ref{S:system} (here we increase $A_{12}$ to 0.08 eV, primarily
increasing the gap at the avoided crossing).
We observe
that both in the position and in the momentum representation, a spurious wave 
packet starts to appear on the lower adiabatic energy level 
as the original wave packet approaches the avoided crossing. Only
when that wave packet starts to vanish again, a much smaller second wave packet
emerges, both at a different position and at a different momentum than 
the first one. The non-adiabatic transition is described by that second 
wave packet, not the first one. While in real world systems (at least for the 
example of NaI below) this effect is much less drastic, it is still 
present to some extent, and it suggests that that 
the adiabatic representation is not the ideal frame of reference for 
understanding transitions at avoided crossings. 

We should also note that only the absolute
values are shown in Figure \ref{fig:stueckelberg}: in reality, all of these
wave packets carry a a rapidly oscillating, nontrivial phase. This makes 
all numerical methods that try to accurately resolve the transmitted wave
packet in the adiabatic representation very expensive, 
as they will need to accurately resolve dynamical 
oscillations that are much larger than the final desired result.

If we are only interested in the crossing probability, i.e.\ the expected 
population on the initially unoccupied adiabatic energy level, 
we can use the Landau-Zener formula, which has been known 
for a long time \cite{Zen32} and also has a firm 
mathematical foundation \cite{JOYE:1991uj}. In a nutshell, the 
Landau-Zener formula first employs a semiclassical approximation to the wave 
packet dynamics, and then avoids resolving the St\"uckelberg oscillations by 
deforming the time axis into the complex plane in the relevant region. 
What the Landau-Zener formula does not provide is information about the 
phase, or more generally any information about the 
transmitted wave packet except 
its size. This is where superadiabatic representations come into play.
 
Just like the {\em adiabatic} approximation improves
on the diabatic one by revealing the 
separation of nuclear dynamics according to electronic energy levels, 
superadiabatic representations improve on the adiabatic one by giving a simpler
dynamical picture in the vicinity of an avoided crossing. Observing the 
wave packet dynamics in higher and higher superadiabatic 
representations initially
reduces the spurious oscillations in the wave packet dynamics, until the
population on the previously unoccupied {\em superadiabatic} 
energy level builds up monotonically as the wave packet travels through 
the avoided crossing: the St\"uckelberg oscillations have disappeared. 
The order of superadiabatic representation where this happens 
is called the optimal one. 
Going to even higher superadiabatic representations from that point on 
reveals the asymptotic nature of the superadiabtatic expansion: 
in those representations the spurious transmitted
wave packet starts to grow again and 
its size eventually diverges as the order of 
superadiabatic approximation goes to infinity. 

The computation of the unitary operators leading to the 
superadiabatic representations, or of the optimal superadiabatic 
representation,
is usually very difficult. However, it has been discovered 
\cite{BGT09,BG1} that it is possible to give an explicit 
formula for the transmitted wave packet in the optimal superadiabatic 
representation without knowing the unitary transformation leading to it. 
By general theory \cite{Teu03}, all superadiabatic representations
agree with the adiabatic one away from an avoided crossing with very good accuracy.
This leads to a straightforward algorithm for efficiently 
computing transitions across avoided crossings, which has been shown to perform 
well in model systems \cite{BG1}. 

The present work investigates the prospects of the superadiabatic theory
for the description of realistic molecular systems. A simple but prototypical
example, the photodissociation of NaI induced by a femtosecond laser pulse, 
is studied. All aspects related to the detailed description of the
non-adiabatic transitions within the superadiabatic representation 
are discussed and addressed. Comparisions with accurate wave packet dynamics 
calculations demonstrate the accuracy of the superadiabatic theory. Furthermore,
different approximations connecting the superadiabatic wave packet propagation 
with semi-classical calculations based on Landau-Zener formulas are discussed and
their accuracy is studied numerically. 

The paper is organized as follows. In section \ref{S:superadiabatic}, 
we outline the theory of superadiabatic representations and explain how they
lead to a rather explicit formula for non-adiabatic transitions. We also describe how
this formula is used for a straightforward numerical scheme for computing 
transitions.
In section \ref{S:system}, we describe the NaI system quantitatively and give the 
numerical details of our algorithm. Finally, we state the results.

\section{Theory}
\label{S:superadiabatic}

\subsection{Superadiabatic approximations}

In practice, the natural starting point for superadiabatic approximations is the
adiabatic one. For explaining the nature of 
superadiabatic representations it is of advantage to start with the diabatic 
representations and recall how it relates to the adiabatic representation. 
We will not discuss the subtle issues of diabatization and existence of a 
diabatic representation \cite{Mead:1982dg}.  
Instead, we just assume that we start with a molecular system with
one nuclear degree of freedom and two electronic states. 
A diabatic representation has the property that there are no derivative couplings, and 
so the Hamiltonian of the system must have the form 
\be \label{eq:hamiltonian}
H = - \frac{\eps^2}{2} \partial_R^2 \bsI + V(R), \text{ with } V(R) = \left( \begin{matrix} 
	V_{11}(R) & V_{12}(R) \\ V_{21}(R) & V_{22}(R).
\end{matrix} \right)
\ee 
Here $\bsI$ is the $2 \times 2$ identity matrix, and $\eps^2$ is the 
'inverse reduced mass'. It will turn out that we need to know the numerical 
value of $\eps$ accurately for computing quantitatively 
correct transition wave packets. For a di-atomic molecule with nuclear 
masses $m_A$ and $m_B$ (measured in atomic mass units), 
we find that in units of [eV  $\rm{\AA}^2$],
\be \label{epsilon}
\eps^2 \approx  4.18 \ast 10^{-3} \frac{m_A + m_B}{m_A m_B}.
\ee 
For typical molecules (like NaI) we obtain values of approximately 
$10^{-4}$ for $\eps^2$.
 
Solutions of the  time dependent 
Schr\"odinger equation 
\be \label{schroed1}
\ii \hbar \partial_t \psi = H \psi
\ee 
are $\bbC^2$-valued functions $\psi$ of the nuclear separation variable 
$R$. 
Since the off-diagonal elements of the potential energy matrix $V$ are 
usually not small, the components $\psi_1(R,t)$ and $\psi_2(R,t)$ 
of the solution display a complicated dynamical behavior: even if $\psi_2(R,t)$ 
is zero at time $t=0$, 
it will become populated by the dynamics for positive times, and evolve via a 
complicated interaction with the first component. Thus the solution in the 
diabatic representation gives very little insight into the physics behind 
equation \eqref{schroed1}.

A conceptually much better description of the system 
\eqref{eq:hamiltonian} is given by adiabatic theory. One way to formulate it is
to say that the solution to \eqref{eq:hamiltonian} clings to the adiabatic 
eigenstates: if $\phi_1(R)$ and $\phi_2(R)$ 
are the (instantaneous) eigenvectors of $V(R)$, then by 
Born-Oppenheimer theory the dynamics of the 
adiabatic components $\phi(R) \cdot \psi(R,t)$ (scalar product in $\bbC^2$) 
will decouple approximately, 
and follow an effective Schr\"odinger equation where the potential is given
by the eigenvalues of $V$.  This is true except near an avoided crossing, where
St\"uckelberg oscillations appear. 

Superadiabatic representations are a systematic way to find improved frames 
of references that give a simpler description of molecular dynamics near 
the avoided crossing. For describing them, we first transform to the 
adiabatic representation: for each $R$, 
let $U_0(R)$ be the unitary $2 \times 2$ matrix that diagonalizes $V(R)$. 
Then the $\bbC^2$-valued adiabatic wave function 
$\psi_0(R,t)  = U_0(R) \psi(R,t) $ is the solution of the 
{\em adiabatic Schr\"odinger equation}
\be \label{schroed}
\ii \hbar \partial_t \psi_0 = H_0 \psi_0,
\ee
where $H_0 = U_0^{-1} H U_0$, or more explicitly  
\be \label{adiab hamil}
H_0 = - \frac{\eps^2}{2} \partial_R^2 I + 
\left( \begin{matrix}
	V_0^+(R) & 
	- \eps \kappa_1(R) (\eps \partial_R) \\ 
	\eps \kappa_1(R) (\eps \partial_R) & 
	V_0^-(R) 
\end{matrix}  
\right) + \caO(\eps^2).
\ee
Here, $\caO(\eps^2)$ signifies that there are further terms with a prefactor 
of $\eps^2$. These are of no importance, since we will see shortly that the 
operator $\eps \partial_R$ is in fact of order one. 
$V^+_0$ and $V_0^-$ are the upper 
(lower) adiabatic energy levels (the eigenvalues of the matrix $V$). 
The coefficients $\pm \kappa_1(R)$ of the first order differential operators 
on the off-diagonal are known as the adiabatic coupling elements and give the size
of the derivative couplings. Overall, transitions are 
of order $\eps$ (thus small),
and the components $\psi_0^+$ and $\psi_0^-$ evolve approximately independently; 
but in the vicinity of an avoided crossing, $\kappa_1(R)$ becomes large, 
and the approximation of independence deteriorates. Non-adiabatic transitions are 
the result. 

Starting from the representation \eqref{adiab hamil}, the idea of the 
first superadiabatic representation is now rather straightforward. We want to 
find another unitary operator $U_1 = U_0 \tilde U_1 $ so that the off-diagonal 
elements in 
\[
H_1 = U_1^{-1} H U_1 = \tilde U_1^{-1} H_0 \tilde U_1
\]
are even smaller than those of $H_0$. 
We can make this statement more precise using powers of $\eps$, but for
this it is necessary to first 
rescale time in such a way that the speed of the nuclei in the new
units is independent of the value of $\eps$. Rescaling time by a 
factor of $\hbar/\eps$ transforms \eqref{schroed} into 
\be \label{schroed2}
\ii \eps \partial_t \psi_0 = H_0 \psi_0,
\ee
while leaving $H_0$ unchanged. In the new time scale, nuclear wave functions
oscillate with a frequency of the order $\eps^{-1}$, so as announced above, 
applying a derivative
coupling of the form $\eps^2 \kappa_1^\pm(R) \partial_R$ 
actually produces a term of order $\eps$ instead of $\eps^2$. 

We can now be more precise about the statement that in the 
first superadiabatic representation, the off-diagonal elements should 
be smaller: we require them to be  (possibly higher order) 
polynomials in $\eps \partial_R$ with a global prefactor of at most $\eps^2$. 

A systematic way to achieve this for all orders of $\eps$ 
has been found in Ref. \onlinecite{BGT09}. There, unitary operators
$U_n$ are constructed such that the $n$-th superadiabatic Hamiltonian 
$H_n = U_n^{-1} H U_n$ has the following properties:
\begin{itemize}
	\item The diagonal elements of $H_n$ are the same as those of $H_0$, up to corrections that are of order $\eps^2$. In other words, the dynamics inside a given electronic state is the adiabatic one. 
	\item The off-diagonal elements of $H_n$ are $(n+1)$-th order polynomials 
	in $\eps \partial_R$ with $R$-dependent coefficients, and carry a global 
	prefactor of $\eps^{n+1}$. 
	\item The off-diagonal elements of $H_{n+1}$ can be constructed from those 
	of $H_k$, $k \leq n$ by solving a set of ordinary differential equations,
	see Proposition 3.3 of Ref. \onlinecite{BGT09}.  
\end{itemize}
Thus the Hamiltonian in the $n$-the superadiabatic representation 
reads, to leading order in $\eps$: 
\be \label{superad hamil}
H_n = - \frac{\eps^2}{2} \partial_R^2 \bsI + \left( 
\begin{matrix}
V_0^+(R) 	& \eps^{n+1} K_{n+1}^+ \\ 
\eps^{n+1} K_{n+1}^- &  V_0^-(R)
\end{matrix}
\right),
\ee
where the {\em $n$-th superadiabatic coupling element} 
$K_{n+1}^\pm$ is a polynomial in $\eps \partial_R$ with coefficients depending
on $R$. The computation of this polynomial is not trivial, e.g.\ since 
$\partial_Q$ and functions of $Q$ do not commute and an order problem needs 
to be solved. This can be done \cite{BGT09} using symbolic calculus and Weyl quantisation. 

The molecular wave function $\psi_n(t) =  U_n \psi(t)$ in the $n$-th
superadiabatic representation in rescaled time is then the solution of the 
of the $n$-th superadiabatic Schr\"odinger equation 
\be \label{supschr}
\ii \eps \partial_t \psi_n = H_n \psi_n. 
\ee
If we choose the initial condition to be confined in one electronic energy state,
\[
\psi_n(0) = 
\left(
\begin{matrix}
	\psi_n^+(0) \\ \psi_n^-(0) 
\end{matrix}
\right)  
= 
\left(
\begin{matrix}
	\phi \\ 0 
\end{matrix}
\right), 
\]
then first order perturbation theory describes the transitions to the 
initially unoccupied energy level:
to leading order in $\eps$, the second component $\psi_n^-(t)$ of 
$\psi_n(t) = U_n \psi(t)$ is given by 
\be \label{superadiabatic perturbation}
\psi_n^-(t) = - \ii \eps^n \int_0^t \e{-(\ii/\eps) (t-s) H^-} 
K_{n+1}^- \e{-(\ii/\eps) s H^+} \phi \, \dd s  ,
\ee
where $H^\pm = -\frac{\eps^2}{2} \partial_R^2 + V_0^\pm$ 
are the adiabatic Hamiltonians for the respective
electronic energy levels. 

At this point we should remember that while $\eps$ is a small number, it is 
fixed for a given molecular system and is not taken to zero. This means 
that we have no guarantee that by switching to higher and higher superadiabatic
representations, the off-diagonal elements of $H_n$ decrease. The reason is that 
the convergence of $\eps^{n+1}$ to zero as $n \to \infty$ is 
offset by a very fast growth of the coefficients of the polynomials
$K_{n+1}$. In Reference \onlinecite{BGT09} it is shown that as functions of $R$, 
these coefficients are maximal at points $R_{\rm c}$ close to 
where the adiabatic energy 
levels exhibit an avoided crossing, and that as functions of $n$ the products
$\eps^n K_n^{\pm}$ first decrease until they become minimal at some 
$n_{\rm opt}$, after which they start to increase and eventually diverge. 
$n_{\rm opt}$ depends on the difference of adiabatic energy levels at $R_c$ 
and can be characterized by the property that the norm of the 
wave function $\psi_n^-(t)$  \eqref{superadiabatic perturbation}
builds up monotonically as the wave packet travels past the avoided crossing 
until it reaches the final value predicted by the Landau-Zener formula. 
In other words $n_{\rm opt}$ is the representation where 
the St\"uckelberg oscillations have disappeared. 

Superadiabatic unitary operators $U_n$
are very complicated objects. While the adiabatic unitary $U_0$ is 
just a rotation of configuration space (and thus easy to understand and 
to implement on a computer), the $U_n$ are
pseudodifferential operators acting on the full wave function $\psi$ 
for $n \geq 1$. The complicated nature of these transformations
is not surprising:  
in the same way as the adiabatic representation clings to the moving frame 
of reference given by the electronic energy levels, the superadiabatic 
representations try to cling to the complicated behavior observed during a 
nonadiabatic transition in order to represent it in a simple form; but then
the dynamical complexity of the transition must be hidden in the 
transformation itself. 

For the practitioner, this will cast serious doubts on the 
practical value of superadiabatic representations. For example,
if equation \eqref{superadiabatic perturbation} is to be of any practical use,
one would first need to transform the initial condition (which will be given in 
the adiabatic representation) to the $n$-th superadiabatic representation,
which is numerically hopeless. Additionally, any statement about the 
time-evolved wave function obtained from \eqref{superadiabatic perturbation}
would need to be transformed back to the adiabatic representation. So,  
equation \eqref{superadiabatic perturbation} describes a possibly simple dynamics
in a very complicated frame of reference, which is useless without a way of 
translating it back to a frame of reference that we can understand.

These objections are valid if we try to understand the adiabatic behavior 
of the molecular wave function at the 
precise time when it travels through the avoided crossing. But often
we are more interested in the wave function at a time when it has already 
left the vicinity of the avoided crossing. In this case, we can make use of 
a convenient property of the superadiabatic representations \cite{BGT09}: 
when the support of a wave packet $\psi$ has no meaningful overlap with the 
region where an avoided crossing is located, the adiabatic representation
and all superadiabatic representations agree with very high accuracy, 
in other words $U_n \psi \approx U_0 \psi$ for such wave functions. 
This enables us to `bypass' the difficulties of the adiabatic representation
during the nonadiabatic transition event in the following way: 
\begin{enumerate}
	\item While the wave packet is still located well away from the avoided
	crossing, we switch from the adiabatic to the optimal superadiabatic 
	representation. These two representations agree, therefore no change to
	the wave packet is made.
	\item We then follow the dynamics of the wave packet across the 
	avoided crossing in the optimal superadiabatic representation. These
	dynamics will be simpler than the adiabatic ones, but we have no easy way 
	of translating them back to the adiabatic representation while the wave 
	packet is located near the avoided crossing. Nevertheless, the wave packet
	in the optimal superadiabatic representation will split up into two 
	wave packets, each on one of the superadiabatic subspaces. 
	\item We follow the dynamics of both of these wave packets until their
	support is well away from the avoided crossing, then switch back to the 
	adiabatic representation. The two representations agree, so no change
	to the wave packet is made.	 
\end{enumerate}
In practice, this means that we can just use equation 
\eqref{superadiabatic perturbation} all the way, 
where $\phi$ is the initial condition in the {\em adiabatic} 
representation. Likewise, $\psi_n^{-}$ is the transmitted wave packet in the 
{\em adiabatic} representation except when its center is very close 
to the avoided crossing.

\subsection{Non-adiabatic transitions}

While we have now established that solving \eqref{supschr} is useful for 
studying the dynamics of non-adiabatic transitions, 
we still have to find an efficient way to actually solve it. 
More precisely, we need to
compute the optimal superadiabatic coupling elements $K_{n_{\rm opt}}^\pm$ 
and the integral \eqref{superadiabatic perturbation}. This 
can be done with the help of asymptotics beyond all orders: it turns out
\cite{BG1,BGT09} that  
$K_{n_{\rm opt}}^\pm$ has a universal, simple description which 
depends on very few parameters of the model. 
We review the main arguments and the result here and refer to the cited 
references for details.

The relevant quantity that completely determines the nonadiabatic transition 
is the difference of the adiabatic energy levels, 
as a function of $R$, in the vicinity of the avoided crossing. Since the
adiabatic energy levels $V_{\pm}$ do not quite cross, 
we can order them so that $V_+(R) > V_-(R)$ for all relevant $R$.  
We define
\be \label{rho}
\rho(R) = \tfrac12  \big(V_+(R) - V_-(R) \big) > 0,
\ee
and observe that when the avoided crossing is at $R=R_c$, 
then by definition $\rho$ has a local minimum there. 

From the work of Berry and Lim \cite{BeLi93} it is possible to derive a 
nonlinear rescaling of the nuclear configuration space in which the 
{\em adiabatic} coupling elements obtain a universal shape.  
We define the {\em natural scale} by 
\be \label{tau}
\tau(R) = 2 \int_{R_{\rm c}}^R \rho(r) \, \dd r,
\ee
and extend the function $\rho$ and $\tau$ into the complex plane. 
By the theory of Stokes lines \cite{JOYE:1991uj}, 
the analytic continuation of $\rho$ 
has a pair of complex conjugate zeroes at locations $R_{\rm cz}$ and 
$R_{\rm cz}^\ast$ close to $R_c$. Let 
\be \label{tau_c}
\tau_{\rm c} = \tau(R_{\rm cz}).
\ee
Near $R = R_{\rm c}$, the adiabatic coupling elements  
are of the universal form
\be \label{universal adiabatic ce}
\kappa_1(R) = \frac{\ii \rho(R)}{3} \Big[ \Big( \frac{1}{\tau(R) - \tau_{\rm c}^\ast}
 - \frac{1}{\tau(R) - \tau_{\rm c}} \Big) + \kappa_{\rm r}(\tau(R)) \Big],
\ee
where the remainder term $\kappa_{\rm r}$ has singularities of order strictly 
less than $1$ at the points $\tau_c=\tau(R_{\rm cz})$ and 
$\tau_c^\ast=\tau(R_{\rm cz}^\ast)$. 

Universality of the optimal superadiabatic coupling elements follows from 
\eqref{universal adiabatic ce} by the Darboux principle \cite{Dingle,BT05-2},
which guarantees that in the recursion for computing the superadiabatic 
representations \cite{BGT09}, the dominant contribution to  
$K_{n+1}^{\pm}$ 
stems from taking derivatives (with respect to $R$) of 
$K_{n}^\pm$. The shape of high derivatives of 
meromorphic functions is dominated by 
the highest order complex singularities nearby \cite{Berry2005do}, which
means that the remainder terms in \eqref{universal adiabatic ce} 
play no role. We define 
\[
h_n(\tau) = \frac{\ii}{(\tau-\tau_{\rm c}^\ast)^n} 
- \frac{\ii}{(\tau-\tau_{\rm c})^n},
\]
and
\[
\kappa_{n}^-(R) = (n-1)! \,  \rho(R) \frac{ \ii^n }{\pi}   h_n(\tau(R)). 
\]
The dominant contribution to $K^-_{n+1}$  is then given \cite{BG1} 
by the fully symmetrized operator
product of $\ii \eps \partial_R$ 
with the multiplication operator $\kappa_{n+1}^-$:
\be \label{opt superad coupling term} 
K_{n+1}^- \phi = \sum_{j=0}^{n+1} \binom{n+1}{j} \Big( \frac{\eps}{2 \ii}
\Big)^j \left( \partial^j\kappa_{n+1}^- \right) 
(- \ii \eps \partial_R)^{n+1-j} \phi.
\ee
Here, $\partial^j \kappa_{n+1}^-$ is the $j$-th derivative of $\kappa_{n+1}$ 
with respect to $R$. 

Formula \eqref{opt superad coupling term} shows that the operator 
$K_{n+1}^-$ is strongly spatially localized: Since $\kappa_{n+1}^-$ and all
its derivatives are rapidly decaying away from $R_{\rm c}$,  
$K_{n+1}^- \phi = 0$ for a wave packet $\phi$ with support not 
overlapping a small vicinity of $R_{\rm c}$. 
While even the adiabatic coupling element
$K_1^-$ exhibits some of this localization, this 
effect becomes much stronger as we increase $n$. 

For the optimal superadiabatic representation, this 
concentration is strongest, and  
non-superadiabatic transitions happen much more quickly than in the non-adiabatic 
ones. In equation \eqref{superadiabatic perturbation}, 
the consequence is that the integral only has to be evaluated for values of 
$s$ that are very close to the time $s_{\rm c}$ where the 
center of the wave packet $\e{-(\ii/\eps) s H^+} \phi$ 
is at $R_{\rm c}$. This is exploited
in Ref.\ \onlinecite{BG1}: since the integration time in 
\eqref{superadiabatic perturbation} is so short, the 
nuclear dynamics {\em on} both of the adiabatic energy surfaces can 
be replaced by quantum dynamics in the linear approximation of the
adiabatic potentials, for which there is an analytic formula. 
The asymptotic form of $\kappa_{n_{\rm opt}}^-$  can be analyzed, 
and then $\psi_n^-$ in \eqref{superadiabatic perturbation} 
can be expressed by an explicit integral formula (see equation (10) 
of reference \onlinecite{BG1}) which is still complicated but no 
longer contains any propagators. It is analyzed further in 
Ref.\ \onlinecite{BG2}.

In many situations, the time it takes the 
wave packet to travel through the crossing 
region is so short that 
a further simplification gives sufficiently good
results: we use free propagation for the adiabatic dynamics near the avoided 
crossing in formula \eqref{superadiabatic perturbation} instead of approximating the 
adiabatic energy levels by linear ones. 
Then another dramatic simplification takes place \cite{BG1}. Let 
$\psi^+_0(R,t_c)$
be the upper adiabatic component of the wave packet, at the time 
$t_{\rm c}$ when its center arrives at $R_{\rm c}$. 
Then for $t > t_{\rm c}$, the expression  
\eqref{superadiabatic perturbation} can be approximated by 
\begin{equation}
	\label{main formula}
	\psi_n(R,t) = \e{-(\ii/\eps) (t-t_c) H^-} \psi^-(R),
\end{equation}
where $\psi^-(R)$ is a wave packet instantaneously created at time 
$t_c$, and having Fourier transform
\begin{equation} 
\widehat \psi^- (k) = -  
\Theta(k^2 - 4 \delta) \frac{v+k}{2 |v|} 
\e{\ii \tau_{\rm c} |k-v| / (2 \delta \eps) } \hat \psi^+_0 (v, t_{\rm c})
\label{main formula2}
\end{equation}
Here, 
$\delta = \rho(R_{\rm c})$ is half the energy gap at the 
avoided crossing, 
and $k$ is the momentum variable. $\Theta$ is the Heaviside function. The 
Fourier transform needs to be done in the correct scale involving 
$\eps$, i.e.\ 
\begin{equation}
\hat \psi(k) = \frac{1}{\sqrt{2 \pi \eps}} \int \e{-(\ii/\eps) k R} 
\psi(R) \, \dd R.
\label{FT}
\end{equation}
Finally, $v = v(k,\delta) = {\rm sgn}(k) \sqrt{k^2 - 4 \delta}$
is the initial momentum that a classical particle would need to have 
to end up with momentum $k$ after falling down a potential energy 
difference of $2 \delta$. The Heaviside function enforces that no 
smaller momenta appear and that $v$ cannot become complex valued. 

A few  comments about formula \eqref{main formula2} are in order:\\
1. The global sign in any formula relating the two adiabatic subspaces must
be indefinite, due to the arbitrariness when choosing the sign 
of the eigenvectors in the adiabatic representation. Here we choose the 
sign to match the given adiabatic representation of our the NaI model below,
in order to compare with exact dynamics. In Reference \onlinecite{BG1}, a different sign was 
used. \\
2. In NaI, the non-adiabatic transition is from the upper to the lower energy
level, and our formula \eqref{main formula2} reflects that. It turns out 
(see in particular the derivation of formula (4.11) in 
Reference \onlinecite{BGT09}), 
that a very similar formula describes the reverse transitions. 
If the wave packet is initially in the lower 
superadiabatic state, the non-adiabatic transition to the upper 
superadiabatic state is given by 
\begin{equation} 
\widehat \psi^+ (k) = - \frac{\tilde v+k}{2 |\tilde v|} 
\e{\ii \tau_{\rm c} |k-\tilde v| / (2 \delta \eps) } \hat \psi^-_0 
(\tilde v, t_{\rm c}),
\label{main formula2+}
\end{equation}
with $\tilde v(k,\delta) = {\rm sgn}(k) \sqrt{k^2 + 4 \delta}$ 
again being the momentum that a classical particle would need to end 
up with momentum $k$ after {\em jumping up} a potential energy of $2 \delta$.
Note that \eqref{main formula2+} predicts that 
energetically forbidden transitions do not happen: the values of 
$\hat \psi^-_0(k)$ with $|k| < 2 \delta$ do not play any role in the 
computation of $\hat \psi^+$.\\
3. Even though formula \eqref{main formula2} describes the evolution of the 
transmitted wave packet in the optimal superadiabatic representation,
it does not depend on the value of $n_{\rm opt}$. This is a 
consequence of the asymptotic universality properties mentioned above.\\
4. Only local information about the adiabatic 
energy levels near the avoided crossing is used: precisely, what is needed is 
the size of the gap $2 \delta$ 
and the quantity $\tau_{\rm c}$ given in equation \eqref{tau_c}.\\
5. Formula \eqref{main formula2} has 
an obvious algorithmic interpretation, which we will give and use 
at the beginning of the next subsection.\\

A very useful way to think about \eqref{main formula2} is to view it as a 
`local in momentum' refinement of the 
classical Landau-Zener formula. For this, assume that $\delta$ is very small,
i.e.\ the crossing of energy levels is very
narrowly avoided. An expansion in $\delta$ then gives
$|v(k)| \approx |k| - 2\delta/|k|$. Thus in \eqref{main formula2}, 
we can write $(v+k)/2v \approx k/|k|$, and $|k-v| \approx 2 \delta / |k|$. 
For an approximate calculation of 
$\tau_{\rm c}$ as given in \eqref{tau_c}, we can 
note that $\rho(R_{\rm c}) = \delta$, and so the zeroes of its analytic 
continuation are very close to the real line. In view of \eqref{tau} an 
expansion in $(R-R_c)$ seems appropriate. 

However, a naive second order expansion of 
$\rho(R)$ around $R_c$ would give the wrong result. The reason is that,
as has been noticed long ago \cite{BeLi93}, the analytic continuation of 
$\rho$ must vanish {\em like a square root} at its complex zeroes. 
The appropriate expansion is thus 
\be \label{rho_approx}
\rho(R) \approx \sqrt{\delta^2 + g(R-R_{\rm c})},
\ee
with smooth $g$ and $g(0) = g'(0) = 0$, and we have to do the second order 
expansion of $g$. This gives 
$\rho(R) \approx \sqrt{\delta^2 + \alpha^2 (R-R_{\rm c})^2}$ 
with $\alpha^2 = \frac12 g''(0)$. 
With this form of $\rho$ 
both $R_{\rm{cz}}$ and $\tau_{\rm c}$ can be computed analytically.
The result is  
\begin{equation} \label{tau_c approx}
\tau_{\rm c} \approx \ii \frac{ \pi \delta^2 }{2 \alpha}.
\end{equation}
The connection with $\rho''(R_{\rm c})$ is made by twice differentiating 
\eqref{rho_approx} and comparing, and we find $\alpha = \sqrt{\delta 
\rho''(R_c)}$. The final result is that for small $\delta$, formula 
\eqref{main formula2} is well approximated by 
\begin{equation}
	\label{approximate main formula}
	\widehat \psi^- (k) = 
	- \frac{k}{|k|} \Theta(k^2 - 4 \delta)
	\e{- \frac{\pi}{2 \eps} \frac{\delta^{3/2}}{|k| (\rho''(R_{\rm c}))^{1/2}}} \hat 
	\psi^+_0 (v, t_{\rm c})
\end{equation}
A very similar formula appears as equation (4) in the paper 
\cite{BelLasTrig} of Belyaev, Lasser and Trigila, where it 	gives the 
Landau-Zener transition rate for single switch surface hopping. The factor
$|k|$ in the denominator of the exponent is present in our formula but not 
in theirs. The reason is that in the formula of Belyaev et al., the second 
derivative of $\rho$ is taken with respect to a point particle traveling 
on the adiabatic surface, while in our formula it is the curvature of the 
surface itself. Thus if we take $|k|$ as the speed of the point particle, 
the additional factor appears by the chain rule. 

For small $\delta$, an application of 
\eqref{approximate main formula} 
can thus be understood as an execution of the following steps:
\begin{enumerate}
	\item decompose the wave function into plane waves of fixed momentum $k$,
	\item perform a momentum shift dictated by energy conservation (this is 
	the significance of the argument $v$ in $\widehat \psi^+$).
	\item compute the single switch surface hopping Landau-Zener probability
	$P_k$ for a point particle with momentum $k$, 
	\item put the fraction $P_k$ of the wave packet at momentum $k$ on the 
	other adiabatic surface. 
	\item reassemble the wave function from the $k$-slices obtained above.
\end{enumerate}
An application of the actual formula \eqref{main formula2} can be understood in 
a similar way, but where in step 3 we apply a more refined transition 
probability which does not rely on $\delta$ being very small. 

There is, however, a very 
significant difference between \eqref{approximate main formula} and a 
surface hopping formula, which comes from the expression $k/|k|$. It 
indicates that the direction in which the original wave packet traverses
the avoided crossing matters and contributes an overall sign to the 
transmitted wave packet. While for single transitions, this is insignificant,
it matters greatly when two of these generated wave packets interfere. In 
Section \ref{sec:results} we will see that for the example of NaI, this is 
indeed the case. 

\subsection{Implementation} \label{S:Implementation}

Here we present a simple algorithm for computing nonadiabatic transitions using formula 
\eqref{main formula}. As we just discussed, there are conceptual similarities 
to surface hopping \cite{BelLasTrig}. When compared to those methods, ours 
has the advantage of preserving phase information of the wave packet. 
Thus, the present method can correctly capture interference effects. 

Our algorithm assumes that we have a way of propagating wave packets on 
uncoupled adiabatic energy levels, and a way to compute the adiabatic 
energy surfaces to reasonable accuracy in special regions, possibly on the fly. 
It then determines transitions between the superadiabatic energy levels as follows:
\begin{enumerate}
	\item We propagate the adiabatic components $\psi_0^{\pm}$ of the 
	wave packet on their respective adiabatic surfaces, with no coupling
	between the adiabatic levels. Any propagator can be used. 
	\item During the evolution, we monitor the distance 
	$h(t) := V_0^+(t) - V_0^-(t)$ of the electronic energy surfaces at the 
	center of all relevant wave packets. 
	\item When a minimum of $h(t)$ is detected for a wave 
	packet, we estimate the size of the expected transition by using the 
	classical Landau-Zener formula. If the estimated size is larger than
	a user-defined threshold, we 
	\begin{enumerate}
	\item Go back to the point in time when the 
	center was at the location $R_{\rm c}$ 
	of the avoided crossing. 
	\item Determine $\delta$ and $\tau_{\rm c}$ from the adiabatic energy
	surfaces.
	\item Put a wave packet according to \eqref{main formula} on the 
	other electronic energy level. 
	\end{enumerate}
\item Go back to step 1.  
\end{enumerate}

This algorithm is very cheap: apart from a pair of 
Fourier transforms that may be necessary for each application of Step
3c), it has the same cost as the propagator used in step 1). 
More importantly, its quality is not compromised when the desired output
is a small quantity. The relative error of the transmitted 
wave packet is equal to the relative error of the single state
propagator, plus systematic errors that reflect the approximate nature
of formula \eqref{main formula}. 

There are two more comments to make about the algorithm. The first 
concerns the calculation of $\tau_c$ given in \eqref{tau_c}. 
At first sight, 
it seems that we need to compute the analytic
continuation of the quantity $\rho$ which may not be known to a very high 
precision in practice. 
Fortunately, since nonadiabatic transitions are going to be 
negligibly small unless the adiabatic energy gap $\delta$ is small, 
we can use the approximation of $\rho$ given in \eqref{rho_approx},
and thus use \eqref{tau_c approx} instead of the true $\tau_{\rm c}$.   
This way, we only need to know the second 
derivatives of the adiabatic energy levels at the point $R_c$ of 
the avoided crossing. Note, however, that the quantity $\pi/(2\eps)$ that 
multiplies our approximation in \eqref{approximate main formula} 
is usually rather large. To make things worse, the transition probability is
obtained by exponentiating, potentially magnifying any errors we make. So 
it is not clear in all cases how good of an approximation 
\eqref{approximate main formula} is. Below, we investigate the situation 
for the example of NaI, and find that the approximate formula is acceptable. 
In other situations, it may be necessary to find better approximations to 
$\rho(R-R_c)$ for good accuracy. On the other hand, 
a sufficiently detailed knowledge of the adiabatic energy levels is 
anyway a theoretical prerequisite to any meaningful prediction
of non-adiabatic transitions. 

The second comment is about slicing the wave function. 
Formula \eqref{main formula} evaluates the initial wave packet
$\psi^+_0(R,t_{\rm c})$ at the time when its centre 
is on the crossing point. In the derivation of that formula, it is assumed 
that $\psi^+_0$ is localized on the semiclassical scale matching the adiabatic 
propagators in \eqref{superadiabatic perturbation}. 
In other words, we need to assume that the width of $\psi^+_0$ 
is not much larger than $\sqrt{\eps}$. 
In practice, this condition may be violated, and in 
fact this is what happens in the case of NaI below. There, we find 
that $\psi^+_0(R,t_{\rm c})$ is significantly different from zero on an interval of length  about 2\AA, or approximately $16 \sqrt{\eps}$. 
Here, a straightforward
application of formula \eqref{main formula2} would result in a poor accuracy. 
The solution is a moderate slicing the
original wave packet. One can e.g.\ use a partition of unity, i.e.\ 
take compactly supported functions
$g_1, \ldots, g_n$ with $\sum_{j=1}^n g_n(R) = 1$ for all $R$, and define
$\psi_{0,j}^+(R) = g_j \psi_0^+(R, t_{\rm c})$. 
The width of each $g_j$ should be around $\sqrt{\eps}$.

Each wave packet $\psi^+_{0,j}$ is then evolved 
on the upper adiabatic surface for the (possibly negative) 
time $t_j$ it takes for its center to reach $R_{\rm c}$, 
where formula \eqref{main formula} is applied to it. This leads to 
a transmitted wave packet $\psi^-_j$ which is then evolved for the time 
$- t_j$ on the lower energy surface. All of these re-evolved 
$\psi^-_j$ are then summed up to produce the transmitted wave packet 
at time $t_{\rm c}$.   

Note that when the $g_j$ are chosen with width of approximately $\sqrt{\eps}$,
Heisenberg's uncertainty relation does {\em not} pose a 
problem with their propagation. One way to see this is to scale out all the
factors of $\eps$ in \eqref{schroed2} with Hamiltonian \eqref{adiab hamil}.
Thus we rescale time by $\eps$ and space by $\eps^2$ and end up with the
adiabatic Schödinger equation $\ii \partial_t \psi(R) = ( - \frac12 
\partial^2_R \psi(R) + V_0^{\pm}(R/\eps)) \psi(R)$, with initial condition
$\psi_{0,j}^{\pm}(R/\eps)$. In the new scale, each slice 
has a width of order one, speed of order one, and needs to be 
propagated until it has travelled a small  distance of order one. So, we can expect that no 
serious broadening of the wave packet takes place. The only precaution we
need to take is that when the $g_j$ have a relatively sharp cutoff, we create
spurious momenta originating from the steep areas of the sliced wave packets
$\psi_{0,j}^+$. But these momenta are very large and thus 
far away from the mean momentum of the incoming wave packet $\psi_0^j$. 
We can therefore 
remove their effect after applying formula \eqref{main formula2} and
resummation of the slices simply by performing a momentum cutoff that removes 
momenta that are too far away from the one dictated by energy conservation.
In the example of NaI, 30 slices and a cutoff procedure produced excellent
agreement with exact calculations. 

Let us finally remark that although equation \eqref{main formula2} 
was derived by switching to the time scale $\hbar/\eps$, i.e.\ by solving
\eqref{schroed2} instead of \eqref{schroed}, the formula itself is 
instantaneous in time. This means that when applying it, there is no need
to change time scales, and the on-level adiabatic propagators 
can be implemented in the time scale involving $\hbar$ if so desired.

\section{System and numerical details} \label{S:system}

As an example, we treat the paradigmatic photo-dissociation 
of NaI \cite{Zew94}. 
The initial wave 
packet is generated by a modulated pump pulse, 
and then travels towards the 
avoided crossing. The description of the pump pulse as well as 
the potential energy surfaces are taken from the work of Engel 
and Metiu \cite{Engel:1989hs}. 
The only difference is that we will work in the adiabatic 
representation, while Engel and Metiu use the diabatic one for 
constructing the initial conditions. 
However, they also use a rapidly
decaying off-diagonal element in the diabatic representation, 
and so the two representations coincide where the initial 
wave packet is created. We will always use [\AA] as the unit of length and [eV] as the unit of energy. 

\begin{figure}
\includegraphics[width = \columnwidth]{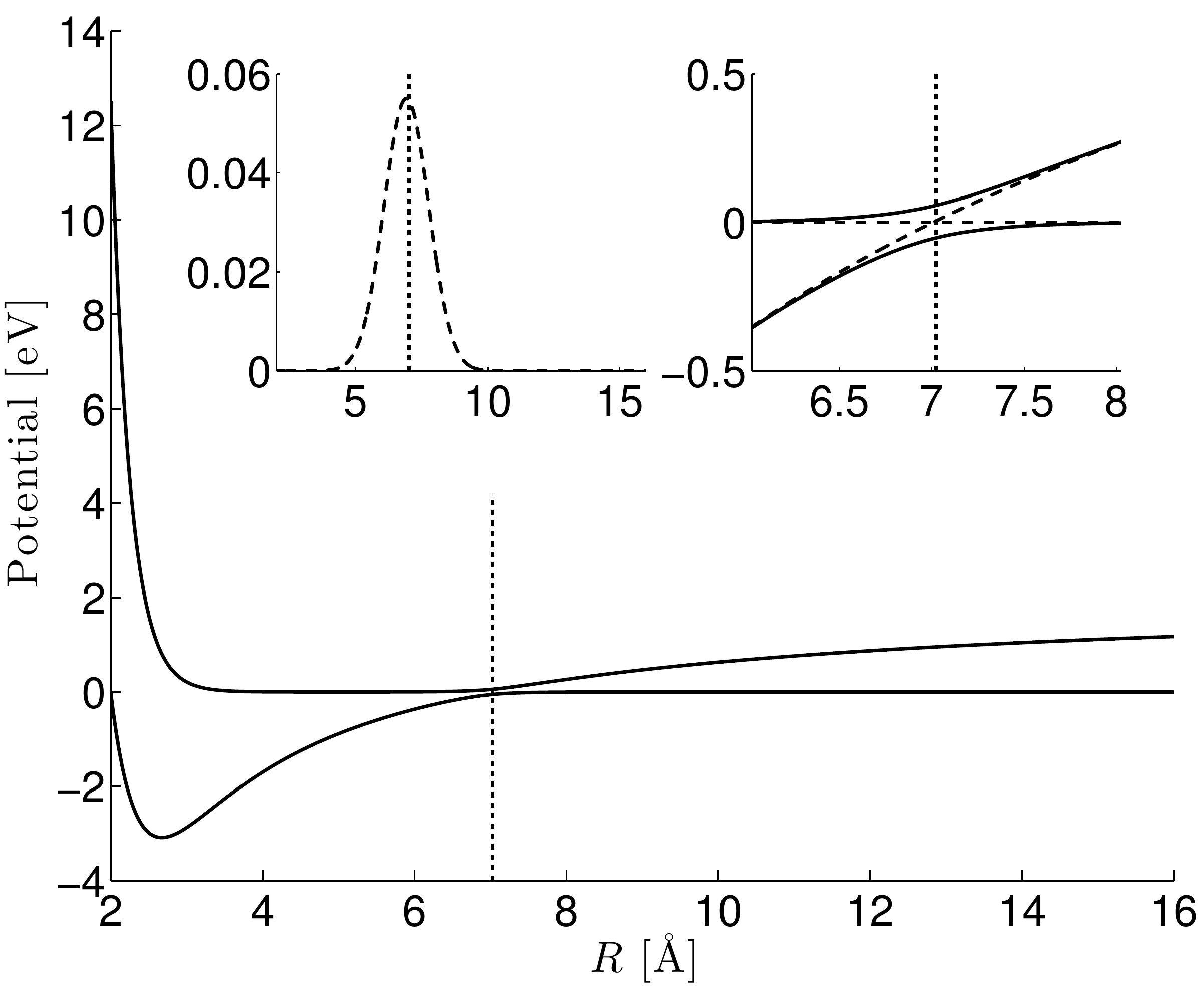}
\caption{Main plot: Adiabatic and diabatic potentials (solid and dashed lines, respectively), which are 
virtually indistinguishable on this scale.  Left inset: The coupling function $V_{12}$.
Right inset: zoom of the main figure around the crossing point, $R_c=7.02$\AA, marked with a 
vertical dotted line in all three plots.}
\label{fig:potentials}
\end{figure}

The Hamiltonian of the model is given by \eqref{eq:hamiltonian}, 
where $|1\rangle$ is the neutral electronic diabatic state, and $|2\rangle$ is the ionic state. 
Engel and Metiu use the ionic potential 
\begin{eqnarray}
\nonumber	V_2(R) := V_{22}(R) & = & (A_2 + (B_2/R)^8) \e{-R/\rho} - e^2/R\\ 
 \nonumber && - e^2(\lambda^+ - \lambda^-) / 
2R^4 - C_2/R^6 \\ && - 2e^2\lambda^+ \lambda^- / R^7 + \Delta E_0
\end{eqnarray}
given in Ref.\ \onlinecite{Faist:1976bn}, and the neutral potential 
\begin{equation}
	V_1(R) := V_{11}(R) = A_1 \exp ( - \beta_1 (R - R_0) )
\end{equation}
from Ref.\ \onlinecite{vanVeen:1981eg}. They choose the diabatic coupling term as 
\begin{equation}
	V_{12}(R) = V_{21}(R) = A_{12} \exp (- \beta_{12} (R-R_x)^2).
\end{equation}
The constants in the above potentials are given in Table \ref{tab1} and 
the potentials,
along with the coupling function, are shown in Figure 
\ref{fig:potentials}. 

Since we need to work with the adiabatic energy surfaces instead of the
diabatic ones, we compute the former from the latter by the formulas
\[
V_0^\pm(R) = \pm \rho(R) + d(R)
\]
with 
\[
\rho = \tfrac12 \sqrt{(V_{11} - V_{22})^2 + V_{12}^2}, \quad 
d = \tfrac12 (V_{11} + V_{22}).
\]
Note also that we obtain $R_c = 7.02 {\rm \AA}$  for the location of the 
avoided crossing in the adiabatic representation, which is slightly
different from $R_x$. 

The atomic masses of Na and I are 23 and 127 atomic mass units, 
respectively, and so \eqref{epsilon} gives  
$\eps^2 \approx 2.147 \ast 10^{-4}$.

\begin{table} \label{tab1}
\caption{Parameters for the potential energy surfaces of NaI (taken from Ref.\ \onlinecite{Engel:1989hs}).}
\begin{tabular}{l l | l l  | l l}
Ionic & & Neutral & & Coupling \\
$A_2$ [eV] & 2760 & $A_1$[eV] & 0.813 & $A_{12}$[eV] & 0.055 \\
$B_2$ [eV$^{1/8}$ \AA] & 2.389 & $\beta_1$ [\AA$^{-1}$] & 4.08 & $\beta_{12}$ 
[\AA$^{-2}$] & 0.6931 \\
$C_2$ [eV \AA$^{6}$] & 11.3 & $R_0$[\AA] & 2.67 & $R_x$[\AA] & 6.93 \\
$\lambda^+$[\AA$^3$] & 0.408 \\
$\lambda^-$[\AA$^3$] & 6.431 \\
$\rho$[\AA] & 0.3489 \\
$\Delta E_0$ [eV] & 0.2075
\end{tabular}
\end{table}

In order to create the initial state on the upper level, the 
excitation via a laser pulse is modeled using a  
first-order perturbation approximation, and in the Condon approximation: 
\cite{Engel:1989hs}
\begin{equation}
	\psi^+(t) = \frac{\ii}{h}\int_0^t \e{-(\ii/\hbar) H^+} E(s)
	                             \exp(- \ii \omega_v s) \phi_v \dd s.
\label{eq:laser1}
\end{equation}
Here $\phi_v$ is the lowest energy eigenstate of the lower level and $\omega_v$ is the 
corresponding vibrational frequency, related to the ground state energy $e_v$ via
$\omega_v = e_v / \hbar$. The ground state wave packet is approximated by a Gaussian.

The electric field of the laser is described by 
\be \label{laser pulse}
	\bsE(s) = \exp(-\ii \omega_0 s) \exp\big[ -\beta (s-s_0)^2 \big].
\ee 
$\omega_0$ is its peak frequency (given by $\omega_0 = 2 \pi c / \lambda$, where $c$ is the 
speed of light and $\lambda$ is the wavelength of the laser). 
As in Reference \onlinecite{Engel:1989hs}, we take  $s_0 = 80$fs and 
$\beta = 1.1 \times 10^{-3}$ fs$^{-2}$ for the pulse width. 
This gives a full width at half maximum of 50fs.

As in reference \onlinecite{Engel:1989hs}, we assume that the laser induced flourescence (LIF) signal is 
proportional to certain populations.  In particular, the LIF signal is assumed to be proportional to the free Na
population, which is taken to be the population of the covalent state to the right of the crossing point
$R_c$,
\begin{equation}
	P_{\rm f} (t) = \int_{R_c}^\infty |\psi_1(R,t)|^2 \dd R.
	\label{eq:Pf}
\end{equation}
Similarly, the bound population is taken to be
\begin{equation}
	P_{\rm b} (t) = \int_0^{R_c} |\psi_1(R,t)|^2 \dd R,
	\label{eq:Pb}
\end{equation}
which measures the population of the covalent state to the left of the crossing point.  It is assumed that the
ionic state population, given by
\begin{equation}
	P_{\rm i} (t) = \int_0^\infty |\psi_2(R,t)|^2 \dd R,
	\label{eq:Pi}
\end{equation}
does not contribute to the LIF signal.  Engel and Metiu \cite{Engel:1989hs} provide a critical analysis of these
definitions.  
In addition, we introduce the adiabatic 
equivalents for the bound
and free populations
\begin{align}
	\tilde{P}_{\rm f} (t) &= \int_{R_{\rm c}}^\infty |\psi_0^- (R,t)|^2 \dd R \label{eq:PfA} \\
	\tilde{P}_{\rm b} (t) &= \int_0^{R_{\rm c}} |\psi_0^+ (R,t)|^2 \dd R \label{eq:PbA},
\end{align}
where $\psi_0^+$ and $\psi_0^-$ are the upper and lower adiabatic populations, respectively. Finally, we define the optimal superadiabatic free population
\be \label{superad free}
\tilde P_{\rm f, sup}(t) = \int_{R_{\rm c}}^\infty | \psi^-(R,t) |^2 \, \dd R,
\ee
where $\psi^-(R,t)$ is computed through formula \eqref{main formula2}. 
Since the coupling $V_{12}$ is localised around $R_x$, for wave packets localised sufficiently
far from the crossing, the definitions of $P_{\rm f/b}$ and  $\tilde{P}_{\rm f/b}$ agree except when
the wave packet is in the crossing region. By superadiabatic theory, 
$\tilde P_{\rm f, sup}(t)$ and $\tilde{P}_{\rm f/b}$ agree 
except when the wave packet is fairly close to the crossing region, 
independently of the shape of $V_{12}$. 
We will see later that for understanding the time evolution 
of the free population, the adiabatic is better than the diabatic one, and the 
optimal superadiabatic one is the best.

The wave packet generated by \eqref{eq:laser1} turns out to be rather 
broad when arriving at the crossing point $R_x$. 
As discussed at the end of Section \ref{S:Implementation}, a slicing procedure is used
to split the wave packet into localized (Gaussian) components. As noted previously, in the present application,
30 slices have been found to be sufficient.

\section{Results} \label{sec:results}

\subsection{Wave packet motion and non-adiabatic transitions}

\begin{figure}
\includegraphics[width = \columnwidth]{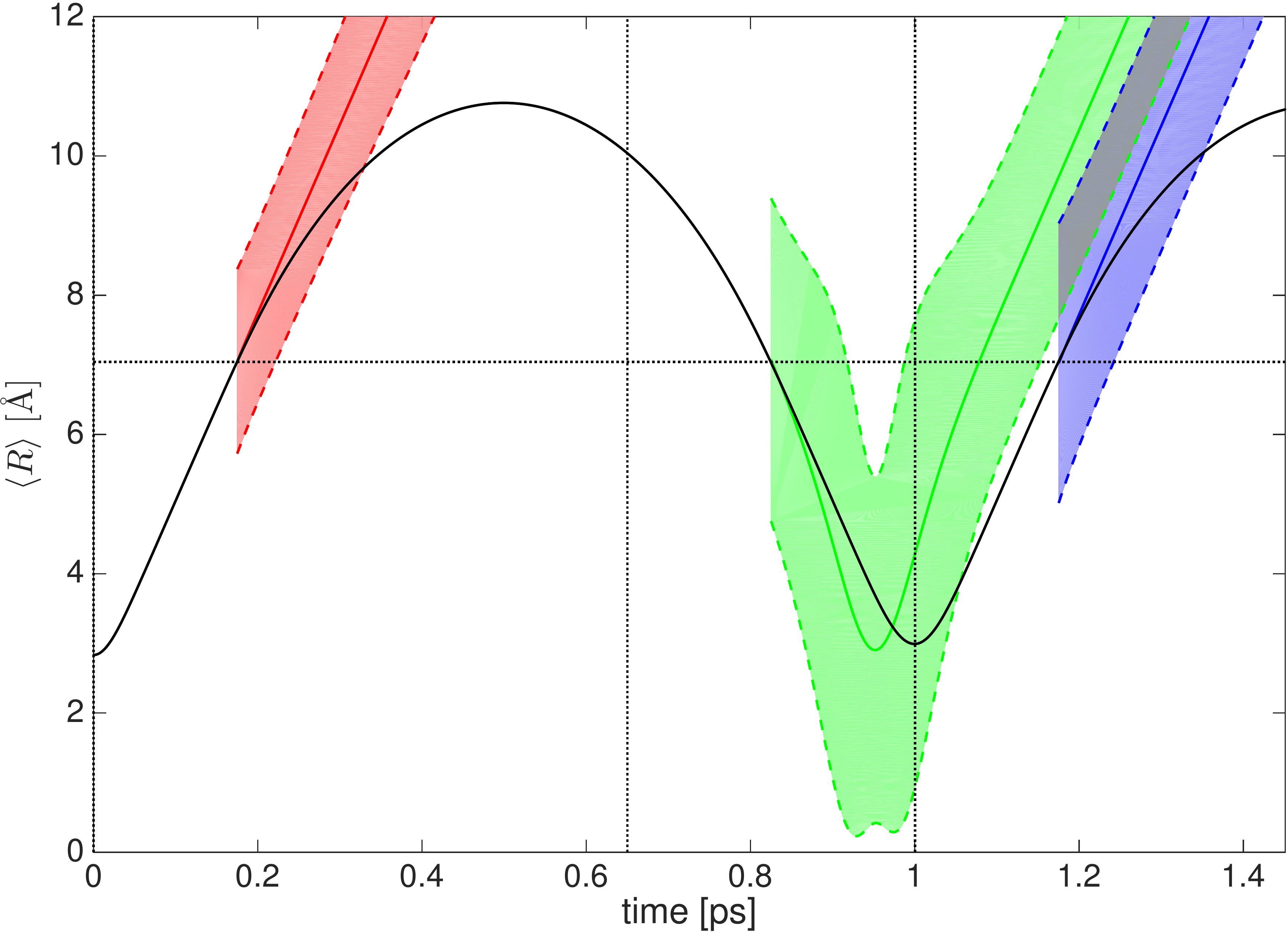}
\caption{In black, the expectation values $\langle R \rangle$
 of the wave packet $\psi^+_0$ as a 
function of time. Transmitted
wave packets are created each time the center of mass crosses $R_c = 7.02 \AA$.
The expectation values $\langle R \rangle$
 of the first, second and third wave packet are shown with 
red, green and blue, solid lines respectively, as functions of time.  We indicate the
spatial delocalization of the transmitted wave packets by shading within 3 standard
deviations of the mean.  Note the grey area denotes the area of significant overlap of
the second and third transmitted wave packets.  Dotted black lines denote $R_c$
and the transmission times.
}
\label{fig:centerOfMass}
\end{figure}

In Figure \ref{fig:centerOfMass} we show the motions of the expectation values $\langle R \rangle$ 
for the various wave packets involved. The wave packet 
generated by the modulated 
laser pulse \eqref{eq:laser1}  travels along the upper adiabatic
surface until it reaches the point $R_{\rm c}$ where the avoided crossing
is located. Here, a wave packet appears on the lower adiabatic surface and 
travels outward. The original wave packet continues to evolve, and after 
being reflected on the right hand side slope of the first excited energy level 
(see Figure \ref{fig:potentials}), it returns to $R_{\rm c}$, where a further 
transmitted wave packet is spawned. Both wave packets are then reflected 
at the left hand side slope of their respective adiabatic energy surface, 
and return to $R_c$ at roughly the same time. A third transmitted wave 
packet is spawned, and creates interference effects with the second one.

\begin{figure}
\caption{Bound (top plot) and free (bottom plot) populations for the diabatic (red, dashed) and
adiabatic (blue, solid) representations, as defined in \eqref{eq:Pf}, \eqref{eq:Pb}, \eqref{eq:PfA}
and \eqref{eq:PbA}, for $\lambda=328$nm. The purple line shows the superadiabatic free population as defined in \eqref{superad free}, and 
generated by instantaneously creating a wave packet according to formula \eqref{main formula}, when the center of $\psi^+_0$ is at 
$R_{\rm c}$.}
\label{fig:328populationsError}
\includegraphics[width = \columnwidth]{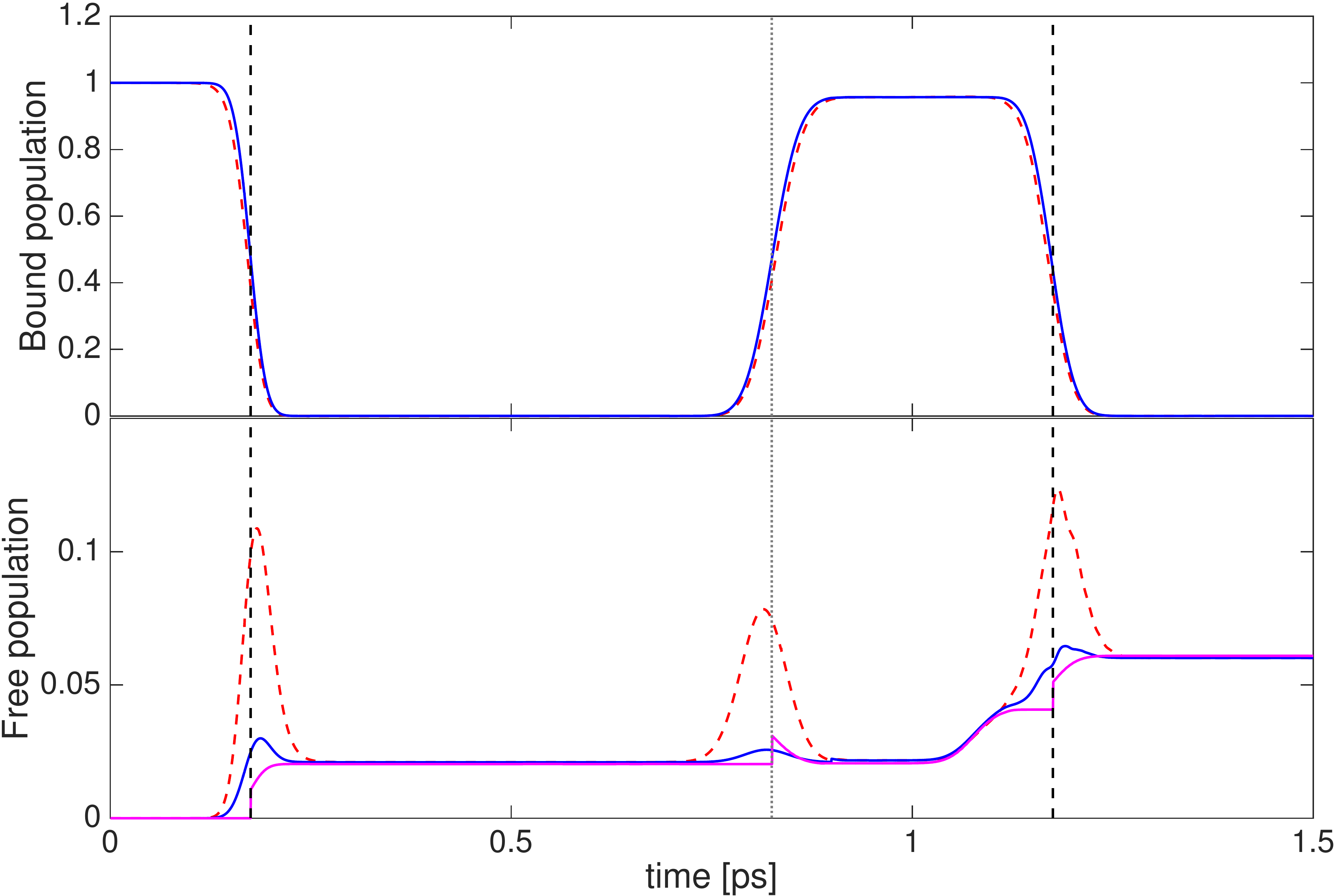}
\end{figure}

Figure \ref{fig:328populationsError} complements Figure \ref{fig:centerOfMass}
by showing the time evolution of the populations 
during the first three visits of the wave packet $\psi^+(t)$ 
to the avoided crossing. 
While the diabatic and adiabatic curves for the bound populations
are almost indistinguishable, the relative size difference 
is significant for the free population. The free population defined via the
diabatic representation (Eq. \eqref{eq:Pf}) 
shows a large spurious maximum whenever the wave packet 
reaches the avoided crossing. This signifies that near the crossing region,
or more generally whenever it does not 
agree with the adiabatic representation, the diabatic 
representation is physically inadequate. 
In the adiabatic representation (blue line), the spurious build-up of 
the transmitted wave packet is already much weaker. It is only about 
30\% larger than the true transmitted wave packet for the first crossing. 
This is an indication that the adiabatic representation is 
rather close to the optimal
superadiabatic representation in the case NaI; in this system, 
St\"uckelberg oscillations are present, but weak. 

The purple line shows the superadiabatic free population 
(Eq. \eqref{superad free}). The discontinuies are artifacts of creating
the transmitted wave packet instantaneously via Eq.\ \eqref{main formula2}.
The subsequent build-up in the first transition signifies that when created, 
the wave packet only half overlaps the region $R > R_x$ 
and subsequently fully enters this
region. The same effect leads to the discontinuity and subsequent die-down of 
the second transition: the wave packet now moves left and leaves the region
$R>R_x$. In the third transition, the first (continuous) build-up of 
free population is due to the return of the lower (super-)adiabatic wave packet 
created in the second transition. A bit later, also the upper adiabatic wave
packet returns to the crossing, and a third transition 
(again with a discontinuity) takes place. This detailed information cannot be 
inferred from the behavior of the adiabatic population at the 
third crossing. In addition, the latter is rather complicated 
due to delicate interference 
effects taking place. We thus see  that the superadiabatic representation
is best suited for understanding the physics of non-adiabatic transitions. 

\subsection{Transmitted wave packet at the first crossing}
\begin{figure}
\caption{Comparison between the result of \eqref{main formula} ($\widehat{\psi^-}$, blue) and transmitted wave packet from the 
full, coupled dynamics ($\widehat{\phi^-}$, black, dotted) at the crossing point
for $\lambda=328$nm in momentum representation.  Top plot shows
the absolute values of the transmitted wave packets in the top panel, and the absolute error in the bottom panel.  
Bottom plot shows the phase and phase error in the top and bottom panels, respectively.
Note that different scales are used to depict the values and their errors.
The inset shows
the wave packet on the upper level at the crossing point.}
\label{fig:328Errors}
\includegraphics[width = \columnwidth]{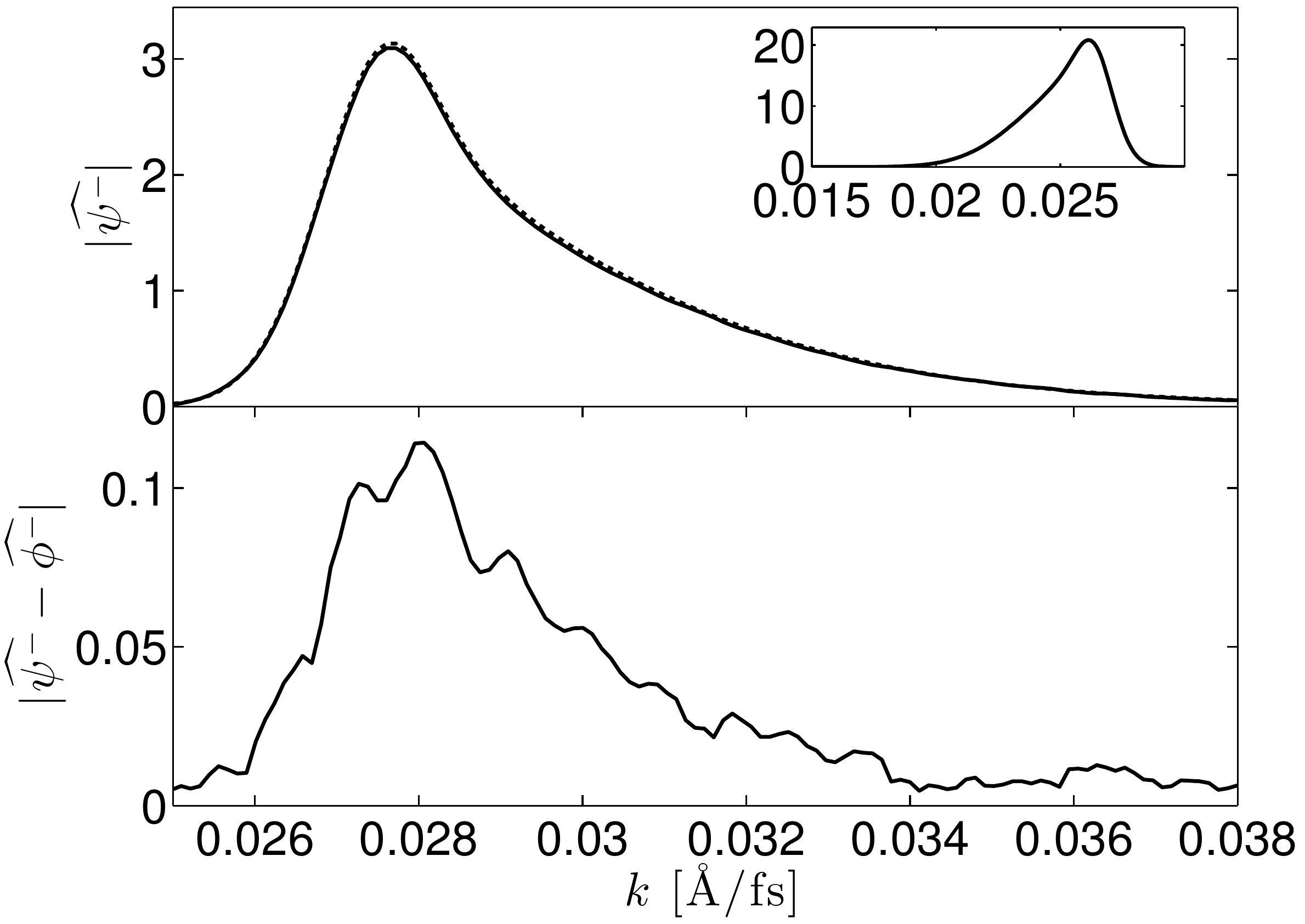}
\includegraphics[width = \columnwidth]{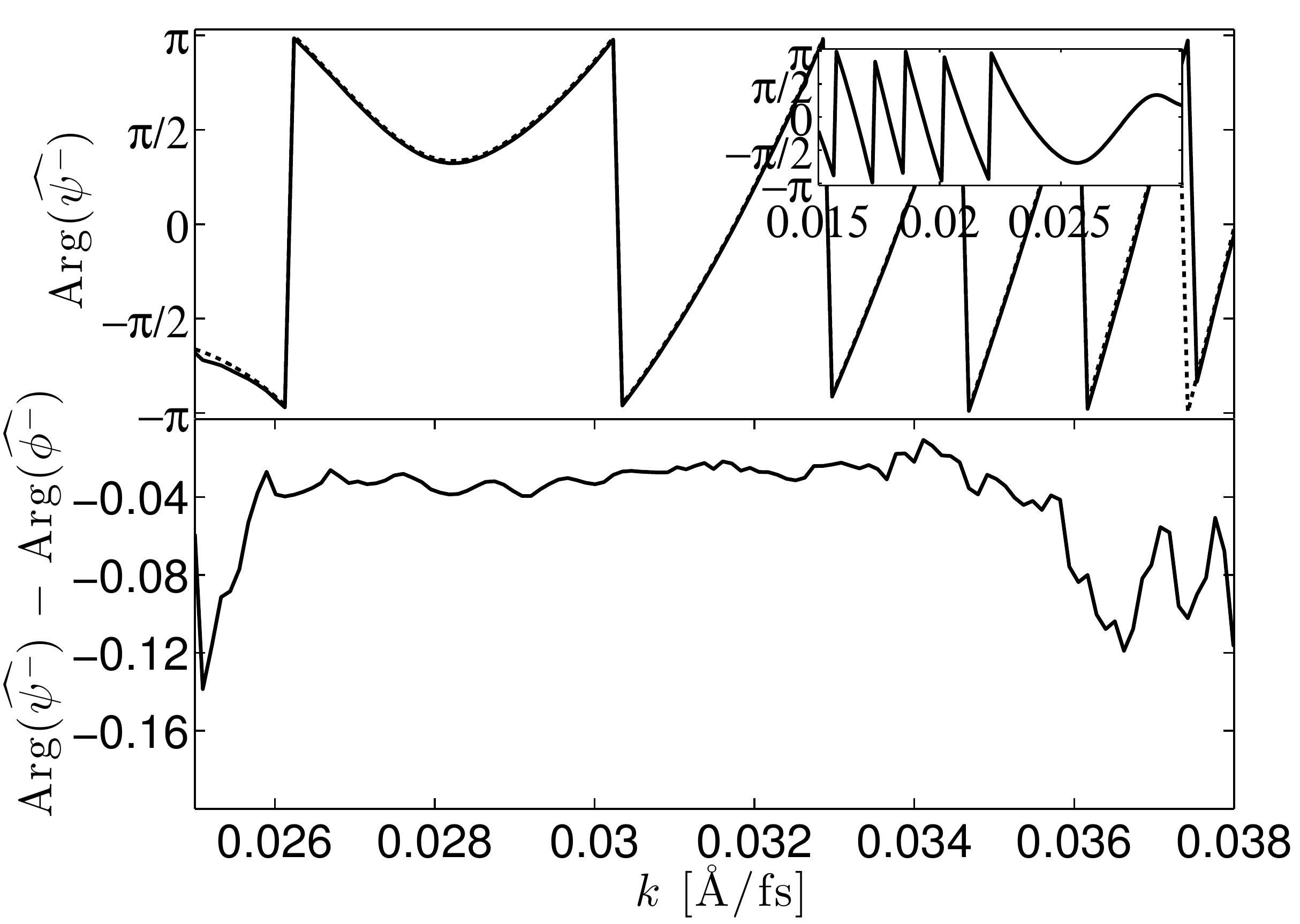}
\end{figure}

Figure~\ref{fig:328Errors} shows the absolute value and phase 
of the transmitted wave packet 
for the first transition with  $\lambda=328$nm.
We plot the wave packets at the crossing point, i.e.\ we compare the results of  \eqref{main formula} 
at $t=t_{\rm c}$ with the wave packet
obtained from running the full, coupled dynamics until the transmitted wave packet is well clear of the crossing region
(in the scattering regime) and then evolving the transmitted wave packet back to the crossing point under the
Born-Oppenheimer approximation.  This is equivalent to evolving the results of \eqref{main formula} into the scattering regime,
but results in a less-rapidly oscillating phase in momentum space.
We plot our results in the momentum representation in order to highlight the 
change of shape that the wave packet $\psi^+_0 $ 
(shown in the inset in momentum representation) undergoes
when making the transition: while the original wave packet has a rather fat
tail of low momenta, these slow parts of the wave packet 
make much smaller non-adiabatic transitions than
the fast ones, and so the transmitted wave packet has instead a rather fat 
tail of high momenta. Note also that neither of the wave functions is 
particularly well approximated by a Gaussian. Also, the phase of both wave 
packets is clearly rather non-trivial. Nevertheless, 
formula \eqref{main formula2} gets it right to very high accuracy. 

The $L^2$ relative error between the results of \eqref{main formula} and the exact calculation is 0.0371
for $\lambda=$328nm.  We also did the calculations for other wavelengths 
of the pump pulse and found that the errors for $\lambda=$300 and 310nm are
0.0240 and 0.0238, respectively.  

\subsection{Combined transmitted wave packets at the second and third
crossing}

We now consider the transmitted wave packet at the avoided crossing point $R_c$ 
at the time $t = 1.18$ ps when the upper adiabatic wave packet $\psi^+$ reaches
$R_c$ for the third time (compare Figure \ref{fig:centerOfMass}). 
The wave packet created by the first visit of 
$\psi^+$ to $R_c$ has long disappeared into the scattering regime, 
but the wave packet created at the second transition now comes back and 
interferes with the one instantaneously created at the third transition.  

\begin{figure}
\caption{Combined transmitted wave packets for second and third transitions for $\lambda=328$nm,
at the third transition time.  Note the relatively good 
agreement between the exact solution (black, dotted), the result of \eqref{main formula} (blue, solid) and 
\eqref{approximate main formula} (red, short dashes).  In contrast, not including the correct prefactor in the LZ formula results in
a wave packet with significant errors (green, dash-dotted).}
\label{fig:328transitions23}
\includegraphics[width = \columnwidth]{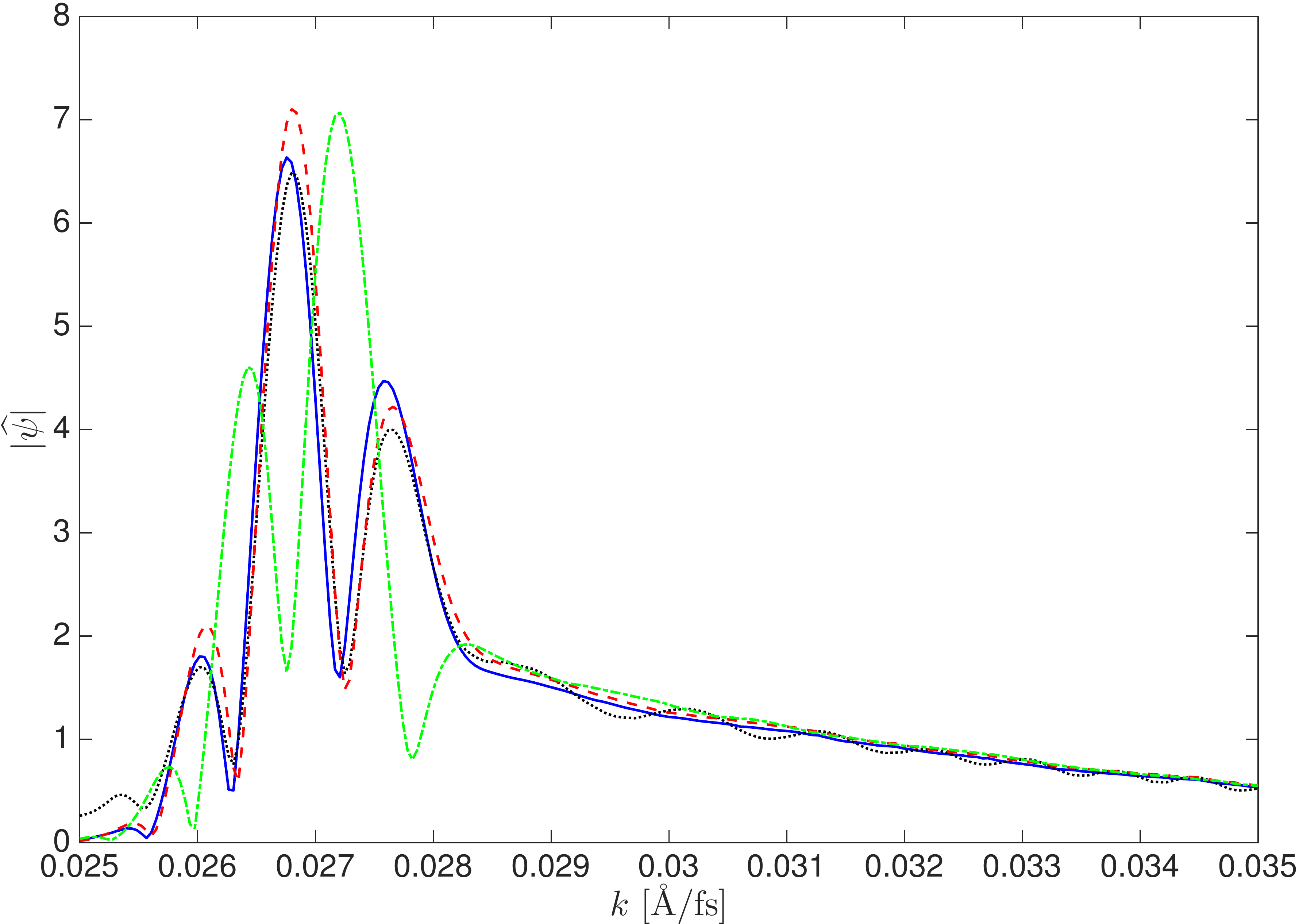}
\end{figure}

Figure \ref{fig:328transitions23} shows the absolute value of the result of this 
interference. The black dotted line is the exact solution, computed by the 
same methodology as for Figure \ref{fig:328Errors}. The blue line is the 
result of applying our algorithm based on \eqref{main formula} 
at the second and third transition time, and adding the result of the 
third transition to the time-evolved result of the second. The resulting 
error is small (7\%). Using the approximate form 
\eqref{approximate main formula} instead of \eqref{main formula2} for 
calculating the transmitted wave packet results in a similar error (6\%, red line). 
This shows that the approximation of very small
$\delta$ is well justified for NaI. Note that the slightly smaller error in 
the second case is a result of a smaller `global' error; the result of 
\eqref{main formula}  is more accurate where the wave packet is large.
The green line, however, indicates 
what happens when we do not take the factor $k/|k|$ into account that 
arises in the limit of small $\delta$ 
from the non-trivial prefactor $(v+k)/2|v|$ found in 
formula \eqref{main formula2}: then the incorrect interference effects lead to 
a prediction that has nothing to do with the true wave packet. 
This pre-factor follows from the optimal superadiabatic theory \cite{BG1} 
and cannot be guessed or obtained by any other means that we know of. 

\subsection{Effect of various approximations on the accuracy}

Towards the end of Section \ref{S:superadiabatic}, we discussed several 
approximations to formula \eqref{main formula2}. Since some of them 
(in particular the approximate calculation of $\tau_{\rm c}$) may be necessary
in cases where we do not have full information about the adiabatic energy
levels, it is interesting to investigate their effect on the quality 
of our algorithm. Here we include a systematic case study of various 
combinations of:\\
(A1) Replacing the non-trivial prefactor $(v+k)/(2|v|)$ with $k/|k| = \pm 1$; \\
(A2) Replacing $\tau_{\rm c}$ 
with the approximation \eqref{tau_c approx}; \\
(A3) Replacing $|k-v|$ with its leading order expansion around $\delta=0$, i.e.\ $2 \delta / |k|$.\\
Using or not using each of these approximations leads to 8 different expressions for the transmitted wave packet, ranging between the full formula
\eqref{main formula} and the Landau-Zener type formula 
\eqref{approximate main formula}.
The $L^2$ and relative errors for each of these approximations (compared to the full formula 
\eqref{main formula}) are given in Table \ref{approxTable} for $\lambda = 328$nm.  The results for 
$\lambda = 300$nm and $\lambda = 310$nm show a similar pattern; the $L^2$ errors
are 0.0240 and 0.0238, respectively for the full formula, whilst the errors when using the 
Landau-Zener approximation are 0.096 and 0.104, respecitvely. Thus while the 
error of about 10\% obtained by using the Landau-Zener type 
formula \eqref{approximate main formula} 
is still acceptable, it is three times larger than the error we get by using the full 
superadiabatic formula \eqref{main formula2}.

\begin{table}
\caption{Error between the wave packet given by \eqref{main formula} with various approximations 
and the exact transmitted wave packet for $\lambda=328$nm.}
\begin{tabular}{rccc|l|l}
& A1 & A2 & A3 & $L^2$ error & relative error to \eqref{main formula}\\
\eqref{main formula} & $\times$ & $\times$ & $\times$ & 0.0371 & 1\\
& $\checkmark$ & $\times$ & $\times$ &  0.0625 & 1.69\\
& $\times$ & $\checkmark$ & $\times$ & 0.1257 & 3.39\\
& $\times$ & $\times$ & $\checkmark$ &0.0665 & 1.79\\
& $\checkmark$ & $\checkmark$ & $\times$ & 0.1286 & 3.47\\
& $\checkmark$ & $\times$ & $\checkmark$ & 0.0401 & 1.09\\
& $\times$ & $\checkmark$ & $\checkmark$ & 0.494 & 4.03\\
\eqref{approximate main formula} & $\checkmark$ & $\checkmark$ & $\checkmark$ &  0.1320 & 3.56
\end{tabular}
\label{approxTable}
\end{table}

\section{Concluding remarks}

Optimal superadiabatic representations are indeed the optimal way to 
describe non-adiabatic transitions at avoided crossings from a 
theoretical point of view. They lead to monotone build up of populations
over time, without spurious populations (St\"uckelberg oscillations) 
appearing at the time of the transition. Formula \eqref{main formula} 
provides a very accurate prediction of the superadiabatic transmitted 
wave packet, which agrees with the adiabatic one away from the crossing 
region. This has been verified in the example of NaI, where in particular
it has been shown that even interference effects at multiple transitions 
are correctly predicted. The algorithm based on superadiabatic representations
can thus provide an inexpensive and accurate way to predict transitions at
avoided crossings, using only local information on the adiabatic energy levels.

The present superadiabatic approach describes a non-adiabatic transition 
as an instanteous transfer process, correctly accounts for phases and 
interference effects, and rests on a solid mathematical basis. It therefore
could provide an interesting starting point for the development 
of semi-classical surface hopping approaches. The connection to the 
Landau-Zener based surface hopping approach of Belayaev, Lasser, and 
Trigila \cite{BelLasTrig}, which has been successfully applied 
to study the non-adiadatic dynamics of NH$_3^+$ \cite{Bel2015}, was outlined.
However, the work of Belayaev et al. is restricted to quasi-classical 
trajectories. It might interesting to combine the superadiabatic description 
of non-adiabatic transitions with, e.g., the semi-classical initial value
\cite{SCIVreview} representation. It could provide an alternative to the 
classical electron analog or mapping approach \cite{mapping1,mapping2} 
frequently used to describe multi-state dynamics in this framework.\\[2cm]

{\bf Acknowledgements:} We would like to thank the Mathematisches 
Forschungsinstitut Oberwolfach and the Banff International Research 
Station for their hospitality during the workshops 1523 
and 16w5006, respectively, where part of this 
research was carried out.

\bibliography{NaI_trans_JCP}

\end{document}